\documentclass[twocolumn,showpacs,preprintnumbers,amsmath,amssymb]{revtex4}

\usepackage{graphicx}
\usepackage{dcolumn}
\usepackage{bm}

\begin{document}

\title{Quantum entanglement and teleportation in quantum dot}
\author{Li-Guo Qin$^{1,2}$}\email{lgqin@shu.edu.cn}
\author{Li-Jun Tian$^{1,2}$}\email{tianlijun@staff.shu.edu.cn}
 \author{Guo-Hong Yang$^{1,2}$}
\affiliation {$^1$Department of Physics, Shanghai University,
Shanghai, 200444, China\\$^2$Shanghai Key Lab for Astrophysics,
Shanghai, 200234, China}
\date{\today}

\begin{abstract}
We study the thermal entanglement and quantum teleportation using
quantum dot as a resource.  We first consider entanglement of the
resource, and then focus on the effects of different parameters on
the teleportation fidelity under different conditions. The critical
temperature of disentanglement is obtained. Based on Bell
measurements in two subspaces, we find the anisotropy measurements
is optimal to the isotropy arising from the entangled eigenstates of
the system in the anisotropy subspace. In addition, it is shown that
the anisotropy transmission fidelity is very high and stable for
quantum dot as quantum channel when the parameters are adjusted. The
possible applications of quantum dot are expected in the quantum
teleportation.
\end{abstract}


\maketitle

\section{Introduction}

Quantum entanglement and quantum teleportation are the fascinating
phenomenons based on the nonlocal property of quantum mechanics, and
play the important role in quantum computation, quantum information
processing and quantum communications. Especially, being one of the
growing interests in quantum information theory, quantum
teleportation has been extensively studied due to teleport unknown
quantum states through the effective quantum channels
\cite{gbs,sbo,idk,mbf,ddb}. At present, quantum teleportation has
received extent investigation both theoretically and experimentally.
For example, as a physical resource, entanglement teleportation via
thermal entangled states of Heisenberg XX \cite{xx}, XY \cite{xy},
XXX \cite{xxx,xxx1}, XXZ \cite{xxz,xxz1} and XYZ \cite{xyz} chain
has been reported, and many optimal schemes based on Bell
measurements are proposed for teleportation \cite{xxx1,sas}.

As the artificial atoms, quantum dot devices provide a
well-controlled object for studying quantum many-body physics.
Ground state-single exciton qubits in quantum dots have been also
proposed for quantum computation architecture \cite{psp}. A
teleportation protocol has been successfully implemented with
photons in the realization of number state qubits \cite{elf}.
Quantum teleportation based on a double quantum dot \cite{kwc,fdp}
and the multielectron quantum dots \cite{ddb} has been studied. In
addition, in vertical dots, a quantitatively new type of Kondo
effect associated with a singlet-triplet degeneracy has been
observed \cite{8}. Furthermore, a generic model of a quantum dot
undergoing the singlet-triplet transition allows for a mapping onto
the impurity Kondo model \cite{mp}. So the characteristics of the
quantum dot are worth investigating. In this paper, we will concern
with the thermal entanglement and teleportation in a quantum dot
from algebra method. Our results will provide experiment with
theoretical foundations on quantum entanglement and teleportation in
quantum dot. From the complicated Hamiltonian of quantum dot, one
can obtain a simple nature Hamiltonian through a effective unitary
matrix. After that, one study effects of the important physical
quantities on the thermal entanglement and teleportation. In the
quantum teleportation process, making use of Bell measurements in
two subspaces, isotropy subspace and anisotropy subspace, one find
that the anisotropy measurements is always optimal to the isotropy
measurements.

Our goal is to study thermal entanglement and quantum teleportation
in a vertical quantum dot with the magnetic field. We consider a
Coulomb-blocked systems and electron-electron interaction to be
relatively weak. It is sufficient to consider two extra electrons in
a quantum dot at the background of a singlet state of all other
$N-2$ electrons, which we will regard as the vacuum. In the case of
even number of electrons $N$ in the dot, these are states with $S=0$
and $S=1$. By finding the effective unitary matrix, one obtain the
nature Hamiltonian of quantum dot in Section II. In Section III, as
the probe of the thermal entanglement, the concurrence $C$ is
studied in the quantum dot. In Section IV, we discuss the quality of
the quantum teleportation using quantum dot in thermal equilibrium
state as a quantum channel. And finally, the conclusions are given.

\section{The Hamiltonian of a quantum dot}
A generic model of a quantum dot can be written from the following
Hamiltonian \cite{11,mpl,mpww}
\begin{eqnarray}
\label{H}
H_{dot}=\sum_{ns}\epsilon_{n}d_{ns}^{+}d_{ns}-E_{s}{\bf{S}_{tot}}^{2}-E_{z}S^{z}+E_{c}(N-N_{0}),
\end{eqnarray}
commuting with the total number of electrons occupying the levels
$n=\pm1$ in the dot, $\hat{N}=\sum_{ns}d_{ns}^{+}d_{ns}$, and with
its total spin,
\begin{eqnarray}
\label{S}
{\hat{\bf{S}}_{tot}}=\frac{1}{2}\sum_{nss^{'}}d_{ns}^{+}\mbox{\boldmath${\bf{\sigma}}$}_{ss^{'}}d_{ns^{'}}.
\end{eqnarray}
is the corresponding total spin of the dot. The operator
$d_{ns}^{+}$ create a electron on a single particle level of the
dot, labeled by the spin $s$ and a discrete quantum number $n$. The
parameters $E_{s}$, $E_{c}$, and $E_{z}=g_{d}\mu_{B}B$ are the
exchange, changing, and Zeeman energies respectively \cite{12} and
$g_{d}$ is the $g$ factor for the electrons in the dot. The
dimensionless gate voltage $N_{0}$ is tuned to an even integer
value. Eq.(\ref{H}) describes the electron-electron interaction at
the mean field level. In general, more complicated interaction terms
should be present in Hamiltonian. These terms are, however,
relatively small for dots with a large number of electrons and
furthermore, they do not influence our discussion of the singlet
triplet transition below, therefore we shall neglect them. For
brevity, we assume that the dot is tuned to the middle of the
Coulomb blockade valley and the level spacing $\delta$ is tunable,
e.g., by means of a magnetic field $B$: $\delta$=$\delta(B)$. If the
level spacing $\delta$ between the last filled and first empty
orbital states happen to be close enough to each other, then the
system will form triplet states to gain energy from the Hund's rule
coupling by rearranging the level occupancy. In this case the ground
state is three-fold degenerate and a Kondo state can be formed. In
order to model the singlet-triplet transition in the ground state of
the dot, it is sufficient to consider these two states. The four
low- energy states of the dot can be labeled as $|S,S^{z}\rangle$ in
terms of the total spin $S=0,1$ and its $z$ projection $S^{z}$
\cite{mpl},
\begin{eqnarray}
\label{1,1}
&&|1,1\rangle={d_{+1\uparrow}^{+}}{d_{-1\uparrow}^{+}}|0\rangle,\nonumber\\
&&|1,0\rangle=\frac{1}{\sqrt{2}}({d_{+1\uparrow}^{+}}{d_{-1\downarrow}^{+}}+
{d_{+1\downarrow}^{+}}{d_{-1\uparrow}^{+}})|0\rangle,\nonumber\\
&&|1,-1\rangle={d_{+1\downarrow}^{+}}{d_{-1\downarrow}^{+}}|0\rangle,\nonumber\\
&&|0,0\rangle={d_{-1\uparrow}^{+}}{d_{-1\downarrow}^{+}}|0\rangle,
\end{eqnarray}
where $|0\rangle$ is the ground state of the dot with $N_{0}-2$
electrons. The transition between the states Eq.(\ref{1,1}) can be
described by the operators
\begin{eqnarray}
\label{S}
{\bf{S}}_{nn^{'}}=\frac{1}{2}\mathcal{P}{\sum_{ss^{'}}}d_{ns}^{+}
\mbox{\boldmath{$\bf{\sigma}$}}_{ss^{'}}d_{n^{'}s^{'}}\mathcal{P},
\end{eqnarray}
where
$\mathcal{P}$=${\sum_{s,s^{z}}}|S,S^{z}\rangle{\langle}S,S^{z}|$ is
the projector onto the ground state manifold (\ref{1,1}). Using the
one-to-one correspondence, we have the following relation between
the states (\ref{1,1}) and the states of two fictitious
$\frac{1}{2}$-spins $\mathbf{S_{1}}$ and $\mathbf{S_{2}}$
\cite{11,mpl}
\begin{eqnarray}
\label{base}
&&|1,1\rangle\Leftrightarrow|\uparrow_1\uparrow_2>,\nonumber\\
&&|1,0\rangle\Leftrightarrow\frac{1}{\sqrt{2}}(|\uparrow_1\downarrow_2>+|\downarrow_1\uparrow_2>),\nonumber\\
&&|1,-1\rangle\Leftrightarrow|\downarrow_1\downarrow_2>,\nonumber\\
&&|0,0\rangle\Leftrightarrow\frac{1}{\sqrt{2}}(|\uparrow_1\downarrow_2>-|\downarrow_1\uparrow_2>).
\end{eqnarray}
By comparing matrix elements directly, there are the following
equations:
\begin{eqnarray}
\label{snn} \mathcal{P}{\sum_{ss^{'}}}d_{ns}^{+}
\mbox{\boldmath$\bf{\sigma}$}_{ss^{'}}d_{ns^{'}}\mathcal{P}={\bf{S}}_{1}+{\bf{S}}_{2},
\end{eqnarray} and
\begin{eqnarray}
\label{dnn}
\mathcal{P}{\sum_{ss^{'}}}d_{ns}^{+}d_{ns^{'}}\mathcal{P}=n[({\bf{S_{1}\cdot\bf{S_{2}}}})-\frac{1}{4}+n].
\end{eqnarray}
After tedious calculations, we find
\begin{equation}
\label{usu} U{\bf S}_1U^{-1}=\frac{1}{2}{\bf \sigma}\otimes I,\ \
\ U{\bf S}_2U^{-1}=\frac{1}{2}I\otimes {\bf \sigma},
\end{equation}
where the unitary matrix and its inverse is respectively
\begin{equation}
U=\left(\begin {array}{cccc} 1&0&0&0\\
0&\frac{\sqrt{2}}{2}&0&\frac{\sqrt{2}}{2}\\
0&\frac{\sqrt{2}}{2}&0&-\frac{\sqrt{2}}{2}\\ 0&0&1&0\end {array}
\right), \ \ \ U^{-1}=\left(\begin {array}{cccc}1&0&0&0\\
0&\frac{\sqrt{2}}{2}&\frac{\sqrt{2}}{2}&0\\ 0&0&0&1 \\
0&\frac{\sqrt{2}}{2}&-\frac{\sqrt{2}}{2}&0 \end {array}\right),
\end{equation}
and ${\bf \sigma}$ and $I$ are Pauli and unit matrix respectively.
So, in the isolated dot-hamiltonian, some operators do not appears,
that is, in some sense they are hidden. They are exposed when
tunnelling between dot and leads is switched on. In terms of Eqs.
(\ref{snn})-(\ref{dnn}), the reduced Hamiltonian of the dot is
written as:
\begin{equation}
\label{h1}
\hat{H}=\frac{k_{0}}{4}\hat{\bf{S}}_1\cdot\hat{\bf{S}}_2-{\gamma}B_{0}
\hat{S}^3.
\end{equation}
Here $\gamma$ is gyromagnetic ratio, and $k_{0}$=$\delta-2E_{s}>0$
is the bare value at $B$=0. $B_{0}$ is the magnetic field of the
degenerate point. We assume that Zeeman energy can be neglected due
to the smallness of the electron $g$ factor, and therefore at the
$B=B_{0}$ point all four states can be considered as degenerate. In
following calculation we will set $\hbar=1$ and the Boltzmann
constant $k=1$. By calculating, the eigenvalues and eigenvectors of
reduced Hamiltonian in Eq. (\ref{h1}) are given by
\begin{eqnarray}
\label{lg2}
&&H|\Psi_{1}\rangle=E_{1}|\Psi_{1}\rangle=(\frac{k_{0}}{16}+\gamma B_0)|00\rangle,\nonumber\\
&&H|\Psi_{2}\rangle=E_{2}|\Psi_{2}\rangle=(\frac{k_{0}}{16}-\gamma
B_0)|11\rangle,\nonumber\\
&&H|\Psi_{3}\rangle= E_{3}|\Psi_{3}\rangle=\frac{k_{0}}{16}[\frac{1}{\sqrt{2}}(|01\rangle+|10\rangle)],\nonumber\\
&&H|\Psi_{4}\rangle=E_{4}|\Psi_{4}\rangle=-\frac{3k_{0}}{16}[\frac{1}{\sqrt{2}}(|01\rangle-|10\rangle)].
\end{eqnarray}
Here $|\Psi_{3}\rangle$ and $|\Psi_{4}\rangle$ are two of Bell
states, which are the maximally entangled states (concurrence
$C=1$). However, $|\Psi_{1}\rangle$ and $|\Psi_{2}\rangle$ are the
disentangled states ($C=0$). It is worth that a threefold degenerate
state of the dot will appear in the absence of the magnetic field.
So the magnetic field just introduces the splitting of energy
levels. $|0\rangle$ stands for spin down and $|1\rangle$ stands for
spin up. These four states are just the singlet and triplet states
in Eq. (\ref{base}).

\section{The thermal entanglement in the quantum dot}

In order to show the entanglement of the quantum dot system, we can
use Wootters concurrence to describe entanglement \cite{shw}
\begin{eqnarray}
\label{lg7} C = max\{\lambda_{1} - \lambda_{2} -\lambda_{3}
-\lambda_{4}, 0\},
\end{eqnarray}
where the parameters $\lambda_{i} (i=1, 2, 3, 4)$ with
$\lambda_{1}\geq\lambda_{2}\geq\lambda_{3}\geq\lambda_{4}$ are the
square roots of the eigenvalues of the operator
\begin{eqnarray}
\label{lg8}
\varsigma=\rho(\sigma^{y}_{1}\otimes\sigma^{y}_{2})\rho^{*}(\sigma^{y}_{1}\otimes\sigma^{y}_{2}).
\end{eqnarray}
Here $\sigma^{y}_{1,2}$ are the Pauli spin matrix of two qubits, and
$\rho$ is the density operator of the system at the thermal
equilibrium, represented by
\begin{eqnarray}
\label{lg122} \label{lg4} \rho= \sum^{4}_{i=1}
p_{i}|\Psi_i\rangle\langle\Psi_i|,
\end{eqnarray}
where $p_i= \exp(-E_i/KT)/Z$ are the probability distributions and
the partition function $Z=Tr[exp(-H/KT)]$. To simplify cumbersome
calculations, we will set the Boltzmann constant
 $K=1$ in following calculation. The concurrence $C$ ranges from $0$ for a separable state to $1$ for a
maximally entangled state. In the standard basis, $\{|11\rangle,
|10\rangle, |01\rangle, |00\rangle\}$, the density matrix $\rho(T)$
of the system reads
\begin{eqnarray}
\label{lg6} \rho(T)=\frac{1}{Z}\left(
           \begin{array}{cccc}
                     u & 0 & 0 & 0 \\
                     0 & w & y & 0 \\
                     0 & y & w & 0 \\
                     0 & 0 & 0 & v \\
                    \end{array}
     \right),
\end{eqnarray}
where the nonzero matrix elements are given by
\begin{eqnarray} \label{lgz}
&&u=exp(-\frac{k_0-16\gamma B_0}{16T}),\nonumber\\
&&w=\frac{1}{2}[exp(-\frac{k_0}{16T})+ exp(\frac{3 k_0}{16T})],\nonumber\\
&&y=\frac{1}{2}[exp(-\frac{k_0}{16T})- exp(\frac{3 k_0}{16T})],\nonumber\\
&&v=exp(-\frac{k_0+16\gamma B_0}{16T}).
\end{eqnarray}
Here $\gamma$ and $B_0$ always appear in the form $\gamma B_0$ as a
whole and thus we can set $\gamma B_0 = r$ in following calculation.
The concurrence has the following form \cite{hf}
\begin{eqnarray}
\label{lgc} C = \frac{2}{Z} max\{|y|-\sqrt{uv},0 \},
\end{eqnarray}
where $Z=Tr[\text{exp}(-H/T)]= u+v+2w$. The system is entangled when
$C> 0$, disentangled for $C=0$ and maximally entangled when $C = 1$.
From Eqs. (\ref{lgz}) and (\ref{lgc}), it is very easy to obtain the
concurrence of this system
\begin{eqnarray}
\label{lgc1} C = max \{\frac{exp(\frac{3k_0}{16T}) - 3
exp(\frac{-k_0}{16T})}{Z},0 \}.
\end{eqnarray}
After calculations, we find that for $k_0 < 0$, the concurrence is
always given by $C=0$. And $C$, as a function of $r$, possesses
$C(r) = C(-r)$, so we will consider only $r>0$ in our calculations.
The influence of parameters on entanglement in quantum dot is
discussed in detail as follows.

\begin{figure}[h]
\includegraphics[angle=0,width=10cm]{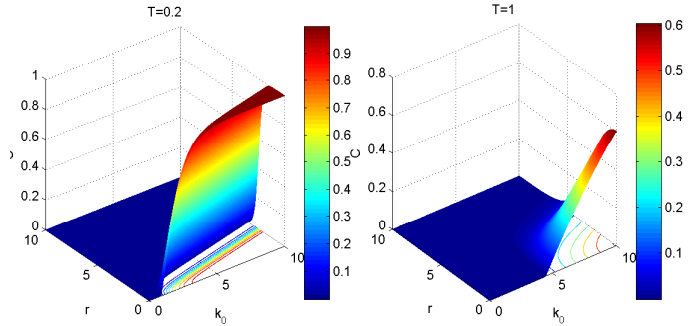}
\caption{(Color online) Concurrence of the quantum dot vs $k_0$ and
$r$ for two different $T$.}
\end{figure}

The concurrences of the quantum dot are ploted in Fig. 1 in terms of
the dimensionless quantities $k_0$ and $r$, for $T=0.2$ and $T=1$.
From Fig. 1 we can see evident differences of the entanglement for
the two cases of different temperature. It is the most obvious that
the region of entanglement becomes smaller with the rise of
temperature. The region of entanglement locates at $r$ smaller, and
$k_0$ bigger. It is clear that $r$ can restrain the entanglement,
however, $k_0$ can enhance the entanglement. Moreover, the maximal
concurrence of quantum dot becomes smaller at $T=0.2$ than $T=1$.
That is to say, the increasing temperature will damage the
entanglement of the system.

\begin{figure}[h]
\includegraphics[angle=0,width=8cm]{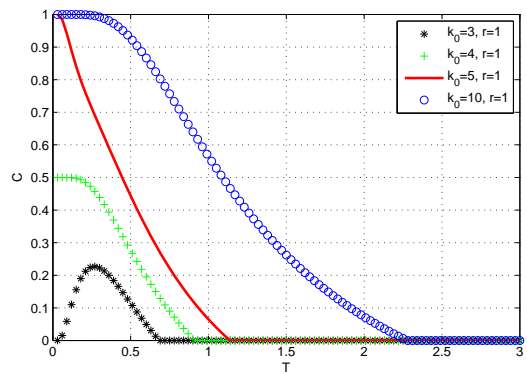}
\caption{(Color online) Concurrence vs temperature ($T$) for
different $k_0$ in the case of $r=1$: $k_0=4$ (green cross line),
$k_0=5$ (red solid line), $k_0=10$ (blue circle line).}
\end{figure}
To better illustrate the relation of the concurrence versus
temperature, we plot the concurrence as functions of the
dimensionless quantities $T$ in the four cases of different $k_0$ at
fixed $r$ in Fig. 2. This figure show $C$ is a monotonically
decreasing function to $T$ with $r \leq \frac{1}{4}k_0$, and all the
curves eventually approach $C=0$(disentanglement), corresponding to
$k_0=4, 5, 10$ and $r=1$. When $r > \frac{1}{4}k_0$, the curve will
firstly raise from $0$ to a top, then decline, finally decrease
$C=0$, corresponding to $k_0=3$ and $r=1$. The reason of them will
be analyzed in the following. The system lies in the ground state at
zero temperature. For $r< \frac{1}{4}k_0$, the ground state is
$|\Psi_{4}\rangle$, which is the maximally entangled state, so $C=1$
for $T=0$. With increasing the temperature, $|\Psi_{4}\rangle$ will
mix with the higher energy levels respectively, namely,
$|\Psi_{2}\rangle$, $|\Psi_{3}\rangle$ and $|\Psi_{1}\rangle$, so
the concurrence monotonically decreases from $1$ to $0$. However,
the ground state become $|\Psi_{2}\rangle$, the disentangled state,
in the case of $r > \frac{1}{4}k_0$, so $C=0$ for $T=0$. Similar the
case, $|\Psi_{2}\rangle$ will mix with $|\Psi_{4}\rangle$,
$|\Psi_{3}\rangle$ and $|\Psi_{1}\rangle$. Hence the concurrence
firstly increases, then decreases. When $r = \frac{1}{4}k_0$, the
ground state will be the degenerate state of $|\Psi_{2}\rangle$ and
$|\Psi_{4}\rangle$. the probabilities of both state in the ground
state are all $50\%$, so $C=0.5$ at $T=0$. These results are all
shown in Fig. 2. In addition, the values of the critical $T_C$, lead
to the system disentanglement $C=0$, can be calculated as
\begin{eqnarray}
\label{lgc2} T_C = \frac{k_0}{4ln3}.
\end{eqnarray}
It shows the relation $T_C$ and $k_0$ is linear and monotonic
increasing. These result shows that $k_0$ can be used as a converter
for $T_C$, be used to adjust the value of $T_C$, namely, change the
temperature of turning on or off the entanglement. So $k_0$ can be a
switch to entanglement, and is tunable, e.g., by means of a external
magnetic field $B$ \cite{mp,mpl}. For this properties, possible
applications are expected in the further.

From the results shown in the above, one may find that $k_0$ may
enhance the concurrence, but $r$ and $T$ may restrain the
entanglement.
\section{Quantum teleportation}

Quantum teleportation via an arbitrary mixed state was first
investigated by G. Bowen and S. Bose \cite{gbs}. They showed when an
arbitrary two-qubit mixed state $\chi$ is used as quantum channel,
the depolarizing (or Pauli) channel is given. Recently F. Caruso et
al. further generalized it to $N$-qubit \cite{fca}.

Now we study the quantum teleportation through the quantum dot using
the standard teleportation protocol $P_0$. Without loss of
generality, we consider the input state is an arbitrary pure state
of a qubit
$|\varphi\rangle_{in}=cos\frac{\theta}{2}|1\rangle+e^{i\phi}sin\frac{\theta}{2}|0\rangle$
($0\leq\theta\leq\pi, 0\leq\phi\leq2\pi$). So the system of the
teleported state and quantum dot in the product state is described
by
\begin{equation}
\label{1.8.1} \rho=\rho_{in}\otimes\rho(T),
\end{equation}
where $\rho_{in}=|\varphi\rangle_{in}\langle\varphi|$ is the density
matrix of the input state. When a joint Bell-basis measurement is
performed on the first two spins, the state of the third spin will
collapse. Under the projection operators $M_{i}$, $\rho$ yields
\cite{xxx1}
\begin{equation}
\label{1.8.2} \rho_{i}=M_i\rho M_i^{\dag},
\end{equation}
where $M_i= E^{i} \otimes I (i=0,1,2,3)$, $E^{0}=
|\Psi^-\rangle\langle\Psi^-|$, $E^{1}=
|\Psi^+\rangle\langle\Psi^+|$, $E^{2}= |\Phi^-\rangle\langle\Phi^-|$
and $E^{3}= |\Phi^+\rangle\langle\Phi^+|$, which
$|\Psi^{\pm}\rangle=\frac{1}{\sqrt{2}}(|10\rangle\pm |01\rangle)$,
$|\Phi^{\pm}\rangle=\frac{1}{\sqrt{2}}(|11\rangle\pm |00\rangle)$.
We can see that $E^{0}$ and $E^{1}$ are in the isotropy subspace and
$E^{2}$ and $E^{3}$ in the anisotropy subspace. By the tracing over
the first two qubits, we can obtain $\rho^{'}_{i} =
\frac{tr_{12}(\rho_{i})}{z}$, which $z=tr(tr_{12}\rho_{i})$. By the
tedious calculations, we can obtain the expressions of
$\rho^{'}_{i}$ respectively
\begin{eqnarray}
\label{1.10.1} &&\rho^{'}_{1}=\frac{1}{z_1}\left(
           \begin{array}{cc}
                     w\cos^2\frac{\theta}{2}+u\sin^2\frac{\theta}{2} & -\frac{1}{2}y\sin\theta e^{-i\delta} \\
                      -\frac{1}{2}y\sin\theta e^{i\delta} & v\cos^2\frac{\theta}{2}+w\sin^2\frac{\theta}{2}  \\
                    \end{array}
     \right),\nonumber\\
    &&\rho^{'}_{2}=\frac{1}{z_1}\left(
           \begin{array}{cc}
                     w\cos^2\frac{\theta}{2}+u\sin^2\frac{\theta}{2} & \frac{1}{2}y\sin\theta e^{-i\delta} \\
                      \frac{1}{2}y\sin\theta e^{i\delta} & v\cos^2\frac{\theta}{2}+w\sin^2\frac{\theta}{2}  \\
                    \end{array}
     \right),\nonumber\\
     &&\rho^{'}_{3}=\frac{1}{z_2}\left(
           \begin{array}{cc}
                     u\cos^2\frac{\theta}{2}+w\sin^2\frac{\theta}{2} & -\frac{1}{2}y\sin\theta e^{i\delta} \\
                      -\frac{1}{2}y\sin\theta e^{-i\delta} & w\cos^2\frac{\theta}{2}+v\sin^2\frac{\theta}{2}  \\
                    \end{array}
     \right),\nonumber\\
     &&\rho^{'}_{4}=\frac{1}{z_2}\left(
           \begin{array}{cc}
                     u\cos^2\frac{\theta}{2}+w\sin^2\frac{\theta}{2} & \frac{1}{2}y\sin\theta e^{i\delta} \\
                      \frac{1}{2}y\sin\theta e^{-i\delta} & w\cos^2\frac{\theta}{2}+v\sin^2\frac{\theta}{2}  \\
                    \end{array}
     \right),\nonumber\\
\end{eqnarray}
where $z_1=w+u\sin^2\frac{\theta}{2}+v\cos^2\frac{\theta}{2}$ and
$z_2=w+v\sin^2\frac{\theta}{2}+u\cos^2\frac{\theta}{2}$. When the
temperature tends to zero, we can note that $\rho^{'}_{1}$ tends to
$\rho_{in}$ arising from $|\Psi_4\rangle=|\Psi^-\rangle$. So at low
temperature, we can obtain the desired teleported state
$\rho^{'}_{1}$ by the $|\Psi^-\rangle$ measurement. In the standard
protocol, the Pauli rotations $\sigma^{j} (j=z,x,y)$ are
respectively applied on $\rho^{'}_{2}$, $\rho^{'}_{3}$ and
$\rho^{'}_{4}$. By the Bell-basis measurement in the isotropy and
anisotropy subspaces, the output states may be given respectively by

\begin{eqnarray}
\label{1.12.1} &&\rho^{e}_{out}=\frac{1}{z_2}\left(
           \begin{array}{cc}
                     w\cos^2\frac{\theta}{2}+v\sin^2\frac{\theta}{2} & -\frac{1}{2}y\sin\theta e^{-i\delta} \\
                      -\frac{1}{2}y\sin\theta e^{i\delta} & u\cos^2\frac{\theta}{2}+w\sin^2\frac{\theta}{2}  \\
                    \end{array}
     \right),\nonumber\\
     &&\rho^{o}_{out}=\rho^{'}_{1}.
\end{eqnarray}
 It is worth noting that $\rho^{e}_{out}=\rho^{o}_{out}$ at $r=0$,
 which is the degenerate point from Eq.(\ref{lg2}). The output states
corresponding to both different subspaces measurement outcomes have
a small difference, which is derived from the magnetic field.

To characterize the quality of the teleported state, the fidelity,
as a useful probe, between $|\varphi\rangle_{in}$ and $\rho_{out}$
is defined by

\begin{eqnarray}
\label{1.12.2} F=\;\;
_{in}\langle\varphi|\rho_{out}|\varphi\rangle_{in}.
\end{eqnarray}
In addition, the average fidelity $F^a$ of teleportation can be
formulated as
\begin{eqnarray}
\label{1.14.1}
F^a=\frac{\int^{2\pi}_{0}d\delta\int^{\pi}_{0}F\sin\theta
d\theta}{4\pi}.
\end{eqnarray}
If quantum dot is used as the quantum channel, making use of  Eqs.
(\ref{1.12.2}) and (\ref{1.14.1}) we can calculate the expressions
of the anisotropy fidelity $F^{o}$, the anisotropy fidelity $F^{e}$
and the average fidelity $F^{a}$.

\begin{figure}[h]
\includegraphics[angle=0,width=9cm]{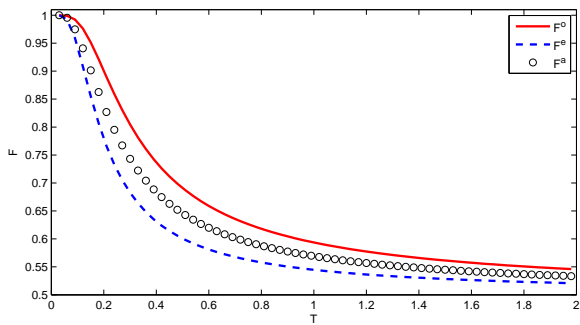}
\caption{(Color online) The fidelity as a function of temperature
$T$ with $k_0=2$ and $r=0.2$. $F^{o}$ and $F^{e}$  are plotted for
$\theta=\frac{\pi}{3}$.}
\end{figure}
The fidelity is plotted as a function of temperature $T$ in Fig. 3.
The evolution of the fidelity, decrease monotonously, is shown as
temperature increases. The evolution curves of $F^{o}$, $F^{e}$ and
$F^{a}$ are very similar in shape, but the value of $F^o$ is always
larger than that of $F^e$. The value of $F^a$ is in the middle of
them. It easily can be seen that three fidelity is equal to one and
and quantum teleportation is perfectly achieved at the zero
temperature, because $|\Psi_4\rangle$ is the ground state and equal
to $|\Psi^-\rangle$ in the conditions of Fig. 3. However, as $T$
increases, not only the thermal entanglement decay, but also the
ground state mix with the excited states, which lead to the fidelity
of teleportation decline. Firstly, the fidelity fall rapidly, then
changes slowly.

\begin{figure}[h]
\includegraphics[angle=0,width=9cm]{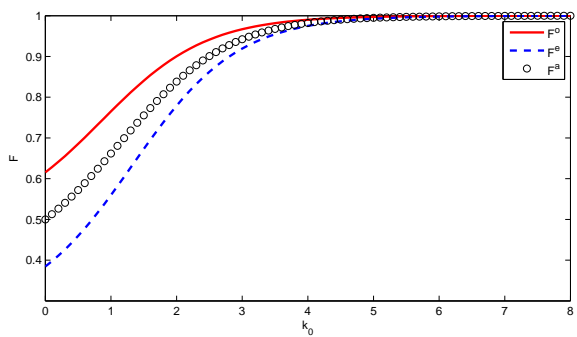}
\caption{(Color online) The fidelity as a function of $k_0$ with
$T=0.2$ and $r=0.2$. $F^{o}$ and $F^{e}$  are plotted for
$\theta=\frac{\pi}{3}$. }
\end{figure}
Figure 4 gives the dependence of the fidelity on $k_0$ at finite
temperature. As $k_0$ increases, the upper results show that the
entanglement increases. Here the fidelity increase monotonically as
$k_0$ increases. For the fixed $T=0.2$ and $r=0.2$, the fidelity
increases rapidly to one and keep perfectly stable. Therefore, $k_0$
is beneficial for the teleportation. Three kinds of the fidelity
always are similar in shape and the value of $F^o$ is always larger
than that of $F^e$.

\begin{figure}[h]
\includegraphics[angle=0,width=9cm]{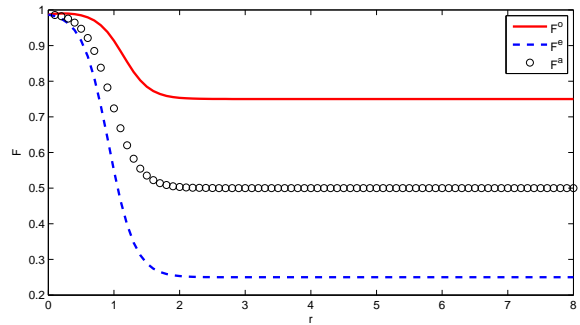}
\caption{(Color online) The fidelity as a function of $r$ with
$T=0.2$ and $k_0=4$. $F^{o}$ and $F^{e}$  are plotted for
$\theta=\frac{\pi}{3}$.  }
\end{figure}
Figure 5 depicts the effects of $r$ on the fidelity of
teleportation. In general, the effect on the quantum teleportation
system is found to be similar. They all first decrease rapidly, then
tend to stable. It is very obvious that $F^o$ is always superior to
$F^e$. When $r=0$, one can see $F^o=F^e=F^a=1$, which indicates that
in this case the quality of the teleportation is perfect. But the
introduction of $r$ not only induces them to separate but also
causes them to decrease in the standard protocol. In addition, the
average fidelity $F^a$ tends steadily to $0.5$. These illustrate
that quantum dot is a better channel for teleportation. These
results show $F^o$ is always superior to $F^e$, because the
eigenstates of the system $|\Psi_{3}\rangle$ and $|\Psi_{4}\rangle$
are just Bell measurements $|\Psi^{+}\rangle$ and $|\Psi^{-}\rangle$
in the anisotropy subspace.

\section{conclusion}
Summarizing, we simplify the Hamiltonian of the quantum dot to the
nature Hamiltonian by integrating and finding the unitary matrix. We
explored how the important physical quantities affect the thermal
entanglement and quantum teleportation. The results show $k_0$ can
improve the entanglement and the quality of the quantum
teleportation. The critical temperature of disentanglement is given.
In addition, we obtain the explicit expression of the output state
of the teleportation based on Bell measurements, with quantum dot as
quantum channel. This allows us to calculate the transmission
fidelity of the quantum channel. Based on Bell measurements in two
subspaces, we found $F^o$ is always optimal to $F^e$ due to arising
from the eigenstates of the system in the anisotropy subspace. It is
shown that the anisotropy transmission fidelity is very high and
stable for quantum dot as quantum channel. These possible
applications are expected in the quantum teleportation.

\section{Acknowledgement}
This work is partly supported by the NSF of China (Grant No.
11075101), Shanghai Leading Academic Discipline Project (Project No.
S30105), and Shanghai Research Foundation (Grant No. 07d222020). The
authors are grateful to Xin-Jian Xu for valuable discussions.


\begin{thebibliography}{99}

\bibitem{gbs}G. Bowen and S. Bose, Phys. Rev. Lett. \textbf{87}, 267901 (2001).
\bibitem{sbo}S. Bose Phys. Rev. Lett. \textbf{91}, 207901 (2003).
\bibitem{idk}I. D. K. Brown, S. Stepney, A. Sudbery, and S. L. Braunstein, J.
Phys. A \textbf{38}, 1119 (2005).
\bibitem{mbf}M. Blasone, F. Dell¡¯Anno, S. De Siena, and F. Illuminati, Phys. Rev. A \textbf{77}, 062304 (2008).
\bibitem{ddb}D. D. B. Rao, S. Ghosh, and P. K. Panigrahi, Phys. Rev. A \textbf{78} 042328 (2008).

\bibitem{xx}Y. Yeo, Phys. Rev. A \textbf{66}, 062312 (2002).

\bibitem{xy}Y. Yeo, T. Q. Liu, Y. E. Lu, and Q. Z. Yang, J. Phys. A \textbf{38}, 3235
(2005).

\bibitem{xxx}G. F. Zhang, Phys. Rev. A, \textbf{75} 034304 (2007).
\bibitem{xxx1}Y. Zhou, G. F. Zhang, S. S. Li, and A. Abliz, Europhys. Lett. \textbf{86} 50004 (2009).
\bibitem{xxz}Y. Zhou, G.F. Zhang, Eur. Phys. J. D, \textbf{47} 227
(2008).
\bibitem{xxz1}J. L. Guo, and H. S. Song, Eur. Phys. J. D, \textbf{56}
265 (2010).
\bibitem{xyz}F. Kheirandish, S. J. Akhtarshenas, and H. Mohammadi, Phys. Rev. A
\textbf{77}, 042309 (2008).
\bibitem{sas}S. Albeverio, S.-M. Fei, and W.-L. Yang, Phys. Rev. A \textbf{66}, 012301 (2002).
\bibitem{psp}P. Solinas, P. Zanardi, N. Zanghi, and F. Rossi, Phys. Rev. A \textbf{67},
052309 (2003).
\bibitem{elf} E. Lombardi, F. Sciarrino, S. Popescu, and F. De Martini, Phys. Rev.
Lett. \textbf{88}, 070402 (2002); Hai-Wong Lee and J. Kim, Phys.
Rev. A \textbf{63}, 012305 (2001)

\bibitem{kwc}K. W. Choo and L. C. Kwek, Phys. Rev. B \textbf{75}, 205321 (2007).
\bibitem{fdp}F. de Pasquale, G. Giorgi, and S. Paganelli, Phys. Rev. Lett. \textbf{93}, 120502 (2004).

\bibitem{8}S. Sasaki, S.De Franceschi, J. M. Elzerman, W. G. van der Wiel,
M. Eto, S. Tarucha, and L. P. Kouwenhoven, Nature \textbf{405}, 764
(2000).
\bibitem{mp}M. Pustilnic, and L. I. Glazman, Phys. Rev. Lett.
\textbf{85} 2993 (2000).

\bibitem{11}M. Pustilnic, and L. I. Glazman, Phys. Rev. Lett.
\textbf{87}, 216601 (2001).

\bibitem{mpl}M. Pustilnik and L. I. Glazman, Phys. Rev. B \textbf{64},
045328 (2001).
\bibitem{mpww}M. Pustilnik, L. I. Glazman, and W.
Hofstetter, Phys. Rev. B \textbf{68}, 161303 (2003).

\bibitem{12}I. L. Kurland, I. L. Aleiner, and B. L. Altshuler,
arXiv:cond-mat/0004205v1.

\bibitem{shw}S. Hill, W. K. Wootters, Phys. Rev. Lett. \textbf{78}, 5022
 (1997); W. K. Wootters, Phys. Rev. Lett. \textbf{80}, 2245 (1998).


\bibitem{hf}H. Fu, A. I. Solomon, and X. Wang, J. Phys. A \textbf{35}, 4293 (2002);


\bibitem{fca}F. Caruso, V. Giovannetti, and G. M. Palma, Phys. Rev. Lett. \textbf{104},
020503 (2010).


\end{thebibliography}
\end{document}